\documentclass[12pt]{amsart}
\usepackage{amssymb}
\usepackage{amscd}
\usepackage{verbatim}
\usepackage{amssymb,amsfonts,pstricks,epsf}
\usepackage[dvips]{graphicx}
\usepackage{epsfig}

\newtheorem{Theorem}{Theorem}[section]

\newtheorem{Lemma}{Lemma}[section]
\newtheorem{Definition}{Definition}[section]
\newtheorem{Corollary}{Corollary}[section]
\newcommand{\thmend}{\hfill \square}

\begin{document}

\def\sR{\hbox{I\kern-.1667em\hbox{R}}}
\newcommand{\R}{\mathbb R}
\newcommand{\M}{\mathcal M}
\newcommand{\m}{{\mathbb M}}
\newcommand{\C}{\mathbb C}
\newcommand{\Z}{\mathbb Z}
\newcommand{\s}{\mathbb S}
\newcommand{\N}{\mathbb N}
\newcommand{\h}{\mathbb H}
\def\rn{{\R}^n}
\def\pair#1#2{\left\langle #1, #2 \right\rangle}
\def\stir#1#2{\left\{ #1, #2 \right\}}
\def\la{\langle}
\def\ra{\rangle}
\def\lap{\Delta}
\def\did{\Delta_{iD}}
\def\orien{{\mathcal O}}
\def\dist{\hbox{dist}}
\def\vol{\hbox{vol}}
\def\spec{\hbox{spec}}
\def\mspec{\hbox{mspec}}
\def\hca{\hbox{hca}}
\def\vp{\text{vp}}
\def\pspec{\hbox{pspec}}
\def\deter{\hbox{det}}
\def\L{{\mathcal L}}
\def\T{{\mathcal T}}
\def\G{{\mathcal G}}
\def\b{{\mathcal B}}

\def\del#1#2{\partial #1/\partial #2}
\def\Del#1#2{\frac{\partial #1}{\partial #2}}
\def\twodel#1#2{\frac{\partial^2 #1}{\partial #2^2}}

\def\pair#1#2{\left\langle #1,#2 \right\rangle}
\def\grad{\nabla}
\def\orienm{w}
\def\whatisthis{N}
\def\sone#1#2{\left\{ \begin{matrix}
                        #1 \\
                        #2  \end{matrix} \right\}}
\def\stwo#1#2{\left[ \begin{matrix}
                        #1 \\
                        #2  \end{matrix} \right]}
\def\choose#1#2{\left( \begin{matrix}
                        #1 \\
                        #2  \end{matrix} \right)}
\def\chone{\hbox{1}}
\def\ddx{\frac{d}{dx}}
\def\ddy{\frac{d}{dy}}


\def\bfa{\text{\rm A}}
\def\bfe{\text{\rm E}}
\def\bfb{\text{\rm B}}
\def\bfc{\text{\rm C}}
\def\bfu{\text{\rm u}}
\def\bft{\text{\rm T}}
\def\fix#1{\text {\rm fix}(#1)}
\def\stab#1{\text {\rm stab}(#1)}
\def\crit#1{\text {\rm crit}(#1)}
\def\osc#1{\text {\rm Osc}(#1)}
\def\rosc#1{\text {\rm R-Osc}(#1)}
\def\signn#1{\text {\rm sign}(#1)}
\def\prob{\text {\rm Prob}}
\def\precur{\text {\rm Prec}}
\def\meas{\text {\rm Meas}}
\def\bv{\text {\rm BV}}
\def\past#1#2{{\text{\rm past}}_{#1}(#2)}
\def\future#1#2{{\text{\rm future}}_{#1}(#2)}
\def\pord{\lll}
\def\rop{{\mathcal R}}
\def\S{{\mathcal S}}
\def\Cs{{\mathcal C}}
\def\tm{{\mathcal M}}
\def\ep{\epsilon}
\newcommand{\St}{{\text {\rm Stab}}(\rop)}

\title{Moment Problems and the Causal Set Approach to Quantum Gravity}

\author{Avner Ash \and Patrick McDonald}

\address{Boston College, Chestnut Hill, MA 02467}

\email{Avner.Ash@bc.edu}

\address{New College of Florida, Sarasota, FL 34243}

\email{ptm@virtu.sar.usf.edu}

\date{August 22, 2002}

\begin{abstract}
We study a collection of discrete Markov chains related to the causal
set approach to modeling discrete theories of quantum gravity.  The
transition probabilities of these chains satisfy a general covariance
principle, a causality principle, and a renormalizability condition.
The corresponding dynamics are completely determined by a sequence of
nonnegative real coupling constants.  Using techniques related to the
classical moment problem, we give a complete description of any such
sequence of coupling constants. We prove a representation theorem:
every discrete theory of quantum gravity arising from causal set
dynamics satisfying covariance, causality and renormalizability
corresponds to a unique probability distribution function on the
nonnegative real numbers, with the coupling constants defining the
theory given by the moments of the distribution.    
\end{abstract}

\keywords{moment problems, causets, covariance, causality,
  renormalization}

\subjclass{44A60, 83F05}

\maketitle

\pagebreak

\baselineskip 24pt

\section{Introduction}\label{se:1}

There are currently a number of approaches aimed at
formulating a successful theory of quantum gravity undergoing
development, the most familiar being String Theory.  This 
note concerns an alternative to String Theory: the {\it Causal Set}
approach to quantum gravity.  In its current state of the
development, the Causal Set approach provides a classical analog to a
true quantum theory; work focussing on  the development of a full
quantum analog is currently underway (cf section 2 below and \cite{RS}
for basic axioms of the Causal Set theory and \cite{R} and \cite{So}
for physical discussions concerning the Causal Set approach).  We
study the Causal Set approach as a classical precursor to a theory of
quantum gravity.   

At first glance, the most natural way to combine quantum theory and
general relativity would be to quantize the spacetime metric.  As is
well known, such a direct approach must contend with a number a
significant obstructions, including the existence of unrenormalizable
divergences.  There is currently no clear consensus as to how these
divergences are to be addressed.  Many believe that the
source of the problem (if not the solution) might lie in the basic
assumptions involving the underlying structure of spacetime.  More
precisely, it has been  suggested that treating spacetime as a
discrete combinatorial object as opposed to a manifold could
lead to insight towards removing the divergences in the quantum field
theoretic approach, if not to a substitute for such an approach (cf
\cite{BLMS}, \cite{RZ} and references therein).

Discrete approaches to gravity initially arose as an attempt to
circumvent many of the difficulties arising in classical general
relativity (eg, existence of singularities, the difficulty of solving
Einstein's field equations for general systems).   Roughly speaking,
the idea behind early discretization procedures involved replacing the
space-time continuum with a triangulation, the construction being
either a triangulation of 4-dimensional spacetime, or later 3+1 
in nature (triangulate a 3-dimensional hypersurface at a fixed time,
triangulate a second hypersurface considered as a time evolution of
the first hypersurface, and connect vertices between triangulated 
hypersurfaces).  In such an approach the vertices are taken to be
(discrete) events, the edges between vertices in different
hypersurfaces spacelike or timelike curves, and the salient relation
between two such events whether one can cause the other or not. 

Over the last two decades, discretization procedures have been
further developed and refined, and their applications in gravity
greatly expanded (cf \cite{RW} for a
recent survey of discrete approaches to gravity, both classical and
quantum).  One particular line of development, pioneered by
Sorkin and his co-workers \cite{BLMS}, de-emphasized the role of the
metric in favor of focussing on the causal structure of
spacetime.  This approach, the so-called Causal Set approach, is
motivated in part by two observations.  First, the causal structure of 
the spacetime continuum determines the topological, differentiable,
and conformal Lorentzian metric structure of the spacetime continuum
(cf \cite{BLMS}, \cite{R}, \cite{RZ}).  Second, the causal structure
of the spacetime continuum and the corresponding discrete
causal structure, are very simple mathematical objects: {\it posets}
(partially ordered sets).  Taking the primary
relationship between two events to be causation, the Causal 
Set approach to gravity posits that the deep structure of spacetime
should be modelled by the discrete causal structures which arise as
natural abstractions of the posets occuring when the
causal structure of the spacetime continuum is discretized (in the
context of gravity, these posets are called {\it causets}).  The Causal
Set approach to gravity then seeks ``natural'' dynamics under which
causets evolve.  In \cite{RS}, Rideout and Sorkin propose such
dynamics (formulated probabilistically)  for the (classical) evolution
of causets.    

Thus, the search for an appropriate dynamical framework for a quantum
theory of gravity has recently led to interest in stochastic
dynamical systems taking their values in certain locally finite
partially ordered sets ({\em causets}).  As discussed in  
\cite{RS}, these systems can be realized as Markov chains whose
transition probabilities are required to satisfy a discrete covariance
principle and a discrete causality principle.  We call such Markov
chains ``generic'' if all of the transition probabilities which could
be nonzero are positive (cf Definition 2.1).  Given the appropriate
mathematical formalism (cf \cite{RS}, section 2 below), it is possible
to classify all such generic chains:  there is a 1-1 correspondence between
generic chains satisfying covariance and causality and nonnegative
sequences of real numbers, $T= \{t_n\}_{n=0}^\infty,$ satisfying
$t_0=1$ (the coupling constants $t_n$ are given explicitly in terms of
the Markov chain - cf \cite{RS} and section 2 below).   

It is easy to see that an arbitrary sequence $T$ is unlikely
to have physical significance, and therefore we want to find
additional natural conditions which restrict the collection of
sequences under consideration to those sequences  which are
``physical.''  Thus, in addition to covariance and causality one might
expect, as first suggested in \cite{MORS}, that a discrete theory of
quantum gravity should satisfy a cosmological renormalizability
condition under cycles of expansion and contraction.  Given the
framework of \cite{RS}, such a condition can be formulated as an
additional constraint on the coupling constants defining the theory.
To make this precise, we introduce the required notation. 

We will denote by $\S$ the collection of sequences of
nonnegative real numbers.  We will denote elements of $\S$ by upper
case roman letters, and, as above, we will use the corresponding lower
case letter to denote specific elements of a given sequence. We will
denote by $\S_1$ the subset of $\S$ consisting of those sequences which
begin with $1.$  We define a {\em cosmological renormalization
  operator} $\rop:\S \to \S$ by   
\begin{eqnarray}
(\rop(T))_n & = t_n + t_{n+1}. \label{rop1}
\end{eqnarray}
The operator $\rop$ admits a stable manifold, $\St \subset \S,$ defined
by 
\begin{eqnarray}
\St & = & \left\{T \in \S: T \in \bigcap_{k=0}^\infty
  \rop^k(\S)\right\}. \label{stab1} 
\end{eqnarray}
We call elements of $\St$ {\em stable sequences} and we note that (cf
\cite{MORS} and section 2 below) there is a 1-1 correspondence 
between generic chains satisfying causality, covariance and cosmological
renormalizability under cycles of expansion and contraction and
elements of $\s = \S_1 \cap \St.$  Our main result, a representation
theorem, gives a complete description of $\St$ in terms of measures on
$\R^+= [0,\infty):$  
\begin{Theorem}\label{maintheorem} Let $T$ be a sequence of 
  nonnegative real numbers.  Then $T\in \St$ if and only
  if there is a nondecreasing function $\alpha:\R^+   \to \R$ such
  that  
\begin{eqnarray}
t_n & = & \int_0^\infty s^n d\alpha(s). \label{moment1}
\end{eqnarray}
\end{Theorem}
For $T$ and $\alpha$ as in Theorem \ref{maintheorem}, we will say that
$T$ is represented by $\alpha.$  

Our theorem is motivated by an
observation of \cite{MORS}: {\em transitive percolation,} the theory
which is determined by choosing $t   \in \R^+$ and defining associated
coupling constants by  
\begin{eqnarray}
t_n & = & t^n, \label{transperc1}
\end{eqnarray}
defines a stable sequence (by convention $0^0=1$).  Transitive
percolation as given by (\ref{transperc1}) is represented by a
probability measure on $\R^+;$ a delta-mass of weight one concentrated
at $t \in \R^+$ has moments which coincide with the sequence.  This
measure can in turn be represented by its probability distribution
function, a translate of the Heaviside function.  Our theorem can be
seen as quantifying to what extent transitive percolation is
representative of the general behavior of stable sequences.  Namely,
any stable sequence is a ``linear combination'' of percolation
sequences. 

As is clear from the statement of Theorem \ref{maintheorem}, our
result is closely related to the classical moment problem of Stieltjes
type (cf section 3 below).  As a consequence, Theorem
\ref{maintheorem} and its proof provide a means of applying the
extensive collection of sophisticated mathematical tools developed in
the context of the moment problem to questions related to quantum
gravity.  We provide a number of straightforward corollaries of our
technique.  These corollaries include an explicit representation of
the transition probabilities associated to any generic Markov chain which
defines a discrete theory satisfying covariance, causality and
cosmological renormalizability, as well as a second representation
theorem which associates to any such theory a natural self-adjoint
nonnegative operator acting on a model Hilbert space (cf section 5
below).    

We thank Rafael Sorkin for suggesting the problem and for a number of
helpful conversations. 

\section{Background and definitions from discrete quantum gravity}

In this section we present the mathematical formulation for a
classical precursor of a discrete theory of quantum gravity.  We
follow the development of \cite{RS} and \cite{MORS}. 

The fundamental object of study, a {\em causet,} is a locally finite
partially ordered set.  Throughout this note we will denote causets
with upper case roman letters and, when needed, indicate the partial
order relation using the symbol $\pord.$  We assume throughout that
$\pord $ is irreflexive.   

An isomorphism of causets is a bijection which preserves the partial
orders.  Isomorphism defines an equivalence relation on causets.  We
will denote by $\Cs_n$ the collection of equivalence classes of causets
with $n$ elements indexed by $\{0,1,\dots, n-1\},$ with partial order
consistent with indexing. Thus, up to equivalence, 
\begin{eqnarray}
\Cs_n & =&  \{C: C \hbox{ a causet }, C = \{a_0, \dots, a_{n-1}\},
a_k\pord a_l \Rightarrow k< l\}.\label{ncausets}
\end{eqnarray}
We write 
\begin{eqnarray}
\Cs&  =&  \bigcup_{n \in \N} \Cs_n \label{causets}
\end{eqnarray}
and we note that $\Cs$ carries a natural partial order given as
follows: $C \prec D$ if and only if $C \in \Cs_n,  \ D\in \Cs_{m}$ with
$n<m,$ and there exists an order preserving function $f:C \to D$ such
that $f(C)$ is an intial segment of $D.$ 

Informally, we can describe the dynamic evolution of causets as
follows: Initially, the state of the system is given by the trivial
causet consisting of a single point.  At each increment of time, an
element comes into existence as the ``offspring'' of elements already
in existence.  That is, at the beginning of the $n$th increment of
time we have a causet $C \in \Cs_n$ which we evolve to a causet $D \in 
\Cs_{n+1}$ by adding an element to $C$ together with relations between
the new element and a subset of elements of $C$ (those elements in the
past of the new element, ie, those which bear some causal relationship
to the new element).  The new relations are determined randomly; the
probability that any given collection of relations is added is given
by a collection of transition probabilities which define the theory.
We can now proceed to formalize this description.

Given a causet $C$ and an element $x \in C,$ we define the past of $x$ by
\begin{eqnarray}
\past{C}{x}&  =&  \{y \in C: y \pord x\}.\label{past}
\end{eqnarray}
We will regard $\past{C}{x}$ as a poset with partial order given by
the partial order of $C.$  A {\em link} in a partially ordered set is
an irreducible relation (ie a relation that contains no other
relation).  A {\em path} in a partially ordered set is a sequence of
elements of the set, each related to the next by a link.   

Given $C \in \Cs_n,$ we will define the family of $C,$ denoted $F(C),$
as those elements $D \in \Cs_{n+1}$ such that $C\prec D,$ where
$\prec$ denotes the partial order of elements of $\Cs:$ 
\begin{eqnarray}
F(C)&  =&  \{D \in \Cs_{n+1}: C \prec D\}.\label{family}
\end{eqnarray}
Given $C \in \Cs_n$ and $D\in F(C),$ the precursor set of the
transition $C \to D,$ denoted $\precur(C,D),$ is the past of the element
$x \in D \setminus C:$  
\begin{eqnarray}
\precur(C,D) & = & \past{D}{x} \subset D. \label{precursor}
\end{eqnarray}
Note that $\precur(C,D)$ is a poset with partial order given by its
description as the past of an element $x\in D.$  The collection of
{\em maximal elements associated to the transition} $C\to D,$ is the
collection of elements of $D$ with links to the element $x:$ 
\begin{eqnarray}
\max(C,D) & = & \{y \in D: y \hbox{ linked to } x, \ x \in D\setminus
C\}.  \label{maximalelements} 
\end{eqnarray}
A special role in the theory will be played by those causets with no
relations. We will denote the element of $\Cs_n$ with no relations by
$A_n:$ 
\begin{eqnarray} A_n &  = &  ((a_0,a_1,
  \dots,a_{n-1}),\emptyset). \label{norelations}
\end{eqnarray}
We note that there is a natural path in $\Cs$ of length $n$ from $A_0$
to $A_n.$  

We define a collection of Markov chains with state space $\Cs$ as follows:

\begin{Definition}\label{mchain1} We say that a Markov chain $M$ with
  state space $\Cs$ belongs to the collection $\M$ if the transition
  probabilities of $M$ satisfy: 
\begin{enumerate}
\item Given $C \in \Cs_n,$ let $\prob(C \to D)$ denote the transition
probability corresponding to an evolution from causet $C$ to causet
$D.$  Then $\prob(C\to D) = 0 $ if $D \notin F(C)$
and $\sum_{D\in F(C)} \prob(C\to D) = 1$
\item \text{(General Covariance)} Let $C \in \Cs_n.$ Suppose ${\mathcal P}_1$ and
${\mathcal P}_2 $ are two paths from $A_0$ to $C$ and write ${\mathcal P}_I =
\{l_{i1},\dots , l_{in}\}$ where the $l_{ij}$ are the links defining the
path ${\mathcal P}_i.$  Then $$\prod_{k=1}^n \prob(l_{1k}) = \prod_{k=1}^n
\prob(l_{2k}).$$
\item \text{(Causality)}  Suppose that $C \in \Cs_n$ and for $i = 1,
  \ 2,$ suppose that $C_i \in F(C).$  Let $B \in \Cs_{m}, \ m \leq n,$
  be defined by 
\begin{eqnarray*}
B & = & \precur(C , C_1) \cup \precur(C , C_2) 
\end{eqnarray*}
with poset structure induced by that of $C.$  Let $B_i \in \Cs_{m+1}$
be $B$ with an element added in the same manner as in the 
transitions $C \to C_i.$  Then we require 
\begin{eqnarray}
\frac{\prob(C\to C_1)}{\prob(C \to C_2)} & = & \frac{\prob(B\to
  B_1)}{\prob(B \to B_2)}. \label{1causality}
\end{eqnarray}
\end{enumerate}
\end{Definition}

It is a theorem of Rideout and Sorkin that any generic element of $\M$ is
completely determined by a discrete collection of coupling constants
given by transitions between causets with no relations.  More
precisely, let $M\in \M,$ and suppose that $A_n $ is given as in
(\ref{norelations}).  Associate to $M$ a sequence of positive coupling
constants $\{q_n\}_{n=0}^\infty$ defined by 
\begin{eqnarray}
q_0 & = & 1 \label{q0} \\
q_n & = & \prob(A_{n-1} \to A_n) \label{qn} 
\end{eqnarray}
where, as above, the expression appearing on the right hand side of
(\ref{qn}) denotes the probability of transition from $A_{n-1} $ to
$A_n.$  In \cite{RS}, Rideout and Sorkin prove that the sequence
$\{q_n\}_{n=0}^\infty $ completely determines the theory associated to
$M.$  More precisely, given an element $C \in \Cs_n,$ and $D \in F(C)$
(cf (\ref{family})), let $\max(C,D)$ be the collection of maximal
elements associated to the transition $C\to D$ (cf
(\ref{maximalelements})) and let $\precur(C,D)$ be the precursor set of
the transition $C\to D$ (cf (\ref{precursor})).
Suppose the cardinality of $\precur(C,D) $ is $\rho$ and that the
cardinality of $\max(C,D)$ is $m. $  Then the transition probability
for the evolution $C \to D$ is given by (cf \cite{RS})
\begin{eqnarray}
\prob(C\to D) & =& q_n \sum_{k=0}^m (-1)^k\choose{m}{k} \frac{1}{q_{\rho-k}}
\label{probcd}
\end{eqnarray}
which indicates that the Markov chain $M$ is completely determined by
the sequence of $q_n$ defined as in (\ref{qn}).  

Following \cite{RS}, we define a sequence $t_n$ by 
\begin{eqnarray}
t_n & = & \sum_{k = 0}^n (-1)^{n-k} \choose{n}{k}
\frac{1}{q_k}. \label{1tn}
\end{eqnarray}
Then we can recover the coupling constants $q_n$ from the sequence of
$t_n:$  
\begin{eqnarray}
\frac{1}{q_n} & = & \sum_{k=0}^n \choose{n}{k} t_k . \label{tkqk}
\end{eqnarray}
As in the introduction, let $\S_1$ be defined by $\S_1 = \{T \in \S:
t_0=1\}.$  There is a bijection between generic elements of $\M$ and elements 
of $\S_1$ given by associating to each element of $\S_1$ the associated 
collection of coupling constants $\{q_n\}_{n=0}^\infty$ given by
(\ref{tkqk}). 

Amongst additional constraints that one might impose to restrict further 
the collection of chains that could serve as classical precursor for a
discrete model of quantum gravity, there is a natural choice involving
cosmological renormalizability under cycles of expansion and
contraction.  More precisely, given a causet $C,$ we call an element
$\gamma \in C$ a {\em post}, if every element of $C$ is either in the
past of $\gamma$ or in the future of $\gamma$ in $C$ (denoted
$\future{C}{\gamma}$): 
\begin{eqnarray}
C & = & \past{C}{\gamma} \cup \{\gamma\} \cup \future{C}{\gamma}.
\end{eqnarray}
Physically, the occurence of a post corresponds to a collapse of the
universe to zero diameter, followed by re-expansion. 

Given a causet $C$ and a post $\gamma,$ there is a simple relationship 
between the coupling constants $t_n$ governing the evolution of $C$
and the coupling constants governing the evolution of the causet
$\{\gamma\} \cup \future{C}{\gamma}$ (cf \cite{MORS}):  If $p =
|\past{C}{\gamma}|,$ then the coupling constants for $\{\gamma\} \cup
\future{C}{\gamma}$ are given by 
\begin{eqnarray}
\tilde{t}_n & = & \sum_{k=0}^p \choose{p}{k} t_{n+k} \label{renorm1}
\end{eqnarray}
where $n>0.$  This relationship is concisely described in terms of the 
cosmological renormalization operator $\rop: \S \to  \S $ defined by 
\begin{eqnarray}
(\rop(T))_n & = & t_n + t_{n+1} \label{renorm2}
\end{eqnarray}
Using the renormalization operator we can write the right hand side of
(\ref{renorm1}) as $(\rop^p(T))_n.$  We use this concise notation to
define the collection of Markov chains which we intend to study.

\begin{Definition}\label{modelsofqg}We say that a Markov chain $M$
  with state space $\Cs$ belongs to the collection $\m$ if 
\begin{enumerate}
\item  $M \in \M$ is generic, and 
\item If $M$ is represented by the sequence $T \in
  \S,$ then 
\begin{eqnarray}
T & \in &  \bigcap_{n=0}^\infty \rop^n(\S). \label{stableset}
\end{eqnarray} 
\end{enumerate}
\end{Definition}
As in the introduction, we call the right-hand-side of
(\ref{stableset}) the stable set of the renormalization operator and
we write  
\begin{eqnarray}
\St & =  &   \bigcap_{n=0}^\infty \rop^n(\S). \label{stableset2} 
\end{eqnarray}  
If we set $\s = \S_1 \cap \St,$ then it is clear from the definition
that there is a bijection between elements of $\m$ and elements of
$\s.$  It is also clear that $\St$ is a convex set.

As discussed in \cite{RS} and \cite{MORS} and our introduction, there
are a number of interesting special cases of processes which satisfy
the conditions defining $\m.$  Of particular interest from our point
of view are theories of {\em transitive percolation} defined by fixing
$t\in \R^+$ and setting 
\begin{eqnarray}
t_n & = & t^n. \label{transitivepercolation}
\end{eqnarray}
As mentioned in the introduction, the sequence defined by
(\ref{transitivepercolation}) can be represented by a 
probability measure on $\R:$ a delta-mass of weight 1 concentrated at
$t \in \R^+.$  This fact, together with the observed convexity of
$\St$ suggests that we develop a representation of $\St$ in terms of
the moments of probability measures on $\R^+.$

\section{Moment Problems}
In this section we develop material related to the classical moment
problem which we will need in the sequel.  References to this material
include \cite{A} and \cite{W}.

Let $[a,b]$ be an interval in the real line, $\alpha:[a,b]
\to \R$ a function of bounded variation.  Given $t \in (a,b), $ we
write 
\begin{eqnarray*}
\alpha(t\pm) & = & \lim_{s\to t^{\pm}}\alpha(s).
\end{eqnarray*}
We say that $\alpha$ is {\em   normalized} if $\alpha(a) = 0$ and for
all $t \in (a,b),$ 
\begin{eqnarray}
\alpha(t) &  =&  \frac{\alpha(t-) +\alpha(t+)}{2}. \label{monotonic21}
\end{eqnarray}
If $f$ is continuous on $[a,b]$ and $\alpha$ is of bounded
variation, we will denote the Stieltjes integral of $f$ with respect to
$\alpha$ by $\int_a^b f(s) d\alpha(s).$  Functions of bounded
variation behave well with respect to Stieltjes integration: if
$\alpha $ is of bounded variation on $[a,b],$ if $f$ is continuous,
and if $c \in [a,b],$ then  
\begin{eqnarray}
\beta(x) & =&  \int_c^x f(s) d\alpha(s) \label{closure1}
\end{eqnarray}
defines a function of bounded variation.  Moreover, if $g$ is
continuous, then
\begin{eqnarray}
\int_a^b g(s) d\beta(s)&  =&  \int_a^b g(s)f(s)
d\alpha(s).\label{closure2} 
\end{eqnarray}
Stieltjes integration behaves as expected under change of coordinates:
if $\alpha$ is of bounded variation on $[a,b],$ if $f$ is continuous
on $[a,b]$ and if $\gamma$ is continuous and strictly increasing on 
$[a,b],$ then 
\begin{eqnarray}
\int_a^b f(s) d\alpha(s)&  =&  \int_c^d f(\gamma(s))
d\alpha(\gamma(s)) \label{change23}
\end{eqnarray}
where $a = \gamma(c)$ and $b = \gamma(d).$  

Stieltjes integration can be extended to improper integrals.  For
example if $\alpha:\R^+ \to \R$ is of bounded variation, and $f$ is
continuous on $(0,\infty),$ we write  
\begin{eqnarray*}
\int_0^\infty f(s) d\alpha(s) & = & \lim_{R\to \infty, \ep \to 0}
\int_\ep^R f(s) d\alpha(s)   
\end{eqnarray*}
when the limit exists and is finite.  The formulas (\ref{closure2})
and (\ref{change23}) are easily extended to improper integrals.  

\begin{Definition}\label{momentproblems}Let $T \in \S.$  We say
  that a nondecreasing function $\alpha :[0,1] \to \R$ is a solution
  of the {\em Hausdorff Moment Problem for $T$} if, for all $n,$ 
\begin{eqnarray}
t_n & = & \int_0^1 s^n d\alpha(s). \label{HMP31}
\end{eqnarray}
We say that a nondecreasing function $\alpha :\R^+ \to \R$ is a
  solution of the {\em Stieltjes Moment Problem for $T$} if, for all $n,$ 
\begin{eqnarray}
t_n & = & \int_0^\infty s^n d\alpha(s). \label{SMP31}
\end{eqnarray}
\end{Definition}
The solution of the Stieltjes Moment Problem played a 
fundamental role in the development of modern analysis.  We recall the
material relevant to our purpose.  
\begin{Definition}\label{derivative}Let $T$ be a sequence of real
  numbers.  The {\em difference operator,} $\Delta,$ mapping
  sequences of real numbers to sequences of real numbers is defined by  
\begin{eqnarray}
(\Delta(T))_n & = & t_{n+1}- t_n. \label{deriv31}
\end{eqnarray}
A sequence $T \in \S$ is said to be {\em completely monotonic}
if for all $n$ and for all $k,$
\begin{eqnarray}
(\Delta^k(T))_n & \geq & 0. \label{cm31}
\end{eqnarray}
\end{Definition}
We can now state Hausdorff's solution to the moment problem bearing
his name: 
\begin{Theorem}\label{HMP32}(Hausdorff) Suppose $T\in \S.$
  Then the Hausdorff Moment Problem for $T$ has a solution if and only
  if the sequence $T$ is completely monotonic.  When $T$ is completely
  monotonic, the solution of the moment problem is unique.  
\end{Theorem}

The solution of the moment problem associated to Stieltjes is given in 
\begin{Theorem}\label{stieltjes} (Stieltjes) Suppose $T\in \S.$ Then
  the Stieltjes Moment Problem for $T$ has a solution if and only if the
  Hankel determinants    
\begin{eqnarray}
H_{0,n} & = & \left| \begin{matrix} 
                       t_0 & t_1 & \dots & t_n \\
                       t_1 & t_2 & \dots & t_{n+1} \\
                       \dots& \dots & \dots & \dots \\
                       t_n & t_{n+1} & \dots & t_{2n} 
                      \end{matrix} \right| \label{hankel1} \\
H_{1,n} & = & \left| \begin{matrix} 
                       t_1 & t_2 & \dots & t_{n+1} \\
                       t_2 & t_3 & \dots & t_{n+2} \\
                       \dots& \dots & \dots & \dots \\
                       t_{n+1} & t_{n+2} & \dots & t_{2n+1} 
                      \end{matrix} \right| \label{hankel2}
\end{eqnarray}
are nonnegative for all values of $n.$
\end{Theorem}

\section{Proof of the main result}

We begin with a definition:
\begin{Definition}\label{tableau} Let $X = (X_{i,j}), \ 0 \leq i,j
  < \infty $ be a doubly infinite matrix with real entries.  We say
  that $X$ is a {\em tableau} if 
\begin{enumerate}
\item $X_{i,j} \geq 0$ for all $i, \ j.$
\item If $X_k = \{X_{k,j}\}_{j=0}^\infty$ is the sequence whose terms
  are given by the $k$th row of $X$ and $\rop$ is the renormalization
  operator defined by (\ref{renorm2}), then $\rop(X_k) = X_{k-1}$ for
  all $k.$  
\end{enumerate}
Given $n\in \N,$ a {\em partial $n$-tableau} is a matrix of $n$ rows
and an infinite number of columns which satisfies the two defining
conditions of a tableau.  If ${\mathcal P}_n$ is the collection of
partial $n$-tableau, if $P \in {\mathcal P}_n$ and $m\leq 
n,$ the $m$-corner operator ${\mathcal O}_m : {\mathcal P}_n \to
\R^m\times \R^m$ is the map defined by truncation:
\begin{eqnarray}
{\mathcal O}_m(P) & = & (P_{i,j}), \ \ \ 0\leq i,j \leq
m-1. \label{mtabs} 
\end{eqnarray}
\end{Definition}
Tableaux are closely related to stable sequences:  It is clear from
Definition \ref{modelsofqg} and Definition \ref{tableau} that if $X$
is a tableau and $X_0 = \{X_{0,n}\}_{n=0}^\infty$ is the first row of
$X,$ then $X_0 \in \St.$  Conversely, 
\begin{Lemma}\label{firstrowstable} Suppose that $T \in \St.$
  Then there is a tableau whose first row is $T.$
\end{Lemma}
{\sc Proof} Let $T \in \St.$  For each $n \in \N$ we can find an
partial $n$-tableau with first row $T.$  We will create an infinite
sequence, $\{Y^\alpha\}_{\alpha =1}^\infty.$  Each $Y^\alpha$ is 
itself an infinite sequence of partial tableaux where the number of
rows will tend to infinity as $\alpha \to \infty.$  Then we will use a
diagonal trick to finish the proof.  

Define a sequence of partial
tableau, $Y^1 = \{Y^1_n\}_{n=1}^\infty,$ where for each $n,$ $Y^1_n$ is a
partial $n$-tableau with $T$ as first row.  Having chosen
subsequences $Y^{m-1} \subset Y^{m-2} \subset \cdots \subset Y^1,$  
choose a subsequence $Y^m$ of $Y^{m-1}$ which satisfies 
\begin{enumerate}
\item  $Y^m_n$ is a partial $k_n$-tableau with $k_n \geq m;$
\item  If ${\mathcal O}_m$ is the $m$-corner operator defined in
  (\ref{mtabs}), then ${\mathcal O}_m(Y_n^m)$ converges as $n \to
  \infty.$   
\end{enumerate}
Consider the sequence of matrices $Z_k = Y_k^k.$  Then $Z_k$ converges
to a doubly infinite matrix with nonnegative entries and first row
given by $T.$  That $Z$ is a tableau follows from the continuity of
the $m$-corner operator acting on $Z_k.$ $\thmend$

\begin{Lemma}\label{columnsmonotonic} Suppose that $X$ is a tableau
  and let $\{X_{k,n}\}= \{X_{k,n}\}_{k=0}^\infty$ be the sequence
  whose terms are given by the $n$th column of $X.$  Then
  $\{X_{k,n}\}$ is a completely monotonic sequence. 
\end{Lemma}

{\sc Proof} An explicit computation shows that the diagonal entries of
$X$ are given by 
\begin{eqnarray}
X_{k,k} & = & \sum_{l=0}^k (-1)^l \choose{k}{l} X_{l,0}.\label{diag41} 
\end{eqnarray}
By assumption the terms of $X$ are all nonnegative.  This proves that the
first column of $X$ is completely monotonic.  To finish the proof,
note that tableau are stable under truncation of their first $n$
columns.   Carrying out such a truncation, the argument above
establishes that the $(n+1)$th column of $X$ (the first column of the
truncated matrix) is completely monotonic. $\thmend$

\begin{Lemma}\label{extending}Suppose $\{y_i\}_{i=1}^\infty$
  is a completely monotonic sequence.  Let $\alpha:[0,1]\to \R$ be the
  normalized nondecreasing function such that, for $1 \leq i,$
\begin{eqnarray}
y_i & = & \int_0^1 s^{i-1} d\alpha(s).\label{y41}
\end{eqnarray}
Then there exists $y_0$ such that $\{y_i\}_{i=0}^\infty$ is completely
monotonic if and only if 
\begin{eqnarray}
  \int_0^1 s^{-1} d\alpha(s) & & \hbox{ converges.}\label{converge}
\end{eqnarray}
Moreover, if $\int_0^1 s^{-1} d\alpha(s) = L,$ then 
\begin{eqnarray}
L &  = &  \inf\{y_0:  \{y_i\}_{i=0}^\infty \hbox{ is completely
  monotonic}\}. \label{lowerboundy0}
\end{eqnarray}
\end{Lemma}
{\sc Proof} Suppose that (\ref{converge}) holds.  Define
$\beta :[0,1] \to \R $ by $$\beta(t) = \int_0^t s^{-1} d\alpha(s).$$
Then $\beta$ is nondecreasing and for $i \geq 1,$ 
\begin{eqnarray*}
\int_0^1 s^i d\beta(s) & = & \int_0^1 s^{i-1} d\alpha(s) 
\end{eqnarray*}
Setting $y_0 = \int_0^1 s^{-1} d\alpha(s),$ we see that there is a
solution to the Hausdorff Moment Problem for the augmented sequence
$\{y_i\}_{i=0}^\infty.$  By Hausdorff's Theorem (cf Theorem
\ref{HMP32}), the augmented sequence is completely monotonic.  

Conversely, suppose there is a $y_0\in \R$ such that the augmented
sequence $\{y_i\}_{i=0}^\infty$ is completely monotonic.  Let
$\beta:[0,1] \to \R$ be the normalized nondecreasing solution of the
Hausdorff Moment Problem for the augmented sequence.  Then, for all $i
\geq 1,$ 
\begin{eqnarray}
\int_0^1 s^{i-1} d\alpha(s) & = & \int_0^1 s^{i-1} sd\beta(s). \label{weier}
\end{eqnarray}
Define continuous linear functionals, $L_\alpha, \ L_\beta,$ on the
space of continuous functions on $[0,1]:$  
\begin{eqnarray*}
L_\alpha(f) & = & \int_0^1 f(s) d\alpha(s) \\
L_\beta(f) & = & \int_0^1 f(s) sd\beta(s) .
\end{eqnarray*}
From (\ref{weier}) we conclude that $L_\alpha$ and $L_\beta$ agree on
polynomials.  By the Weierstrass theorem and continuity of the
integral, we conclude that $L_\alpha = L_\beta.$  Choose $f_n(s)$ the 
increasing sequence of nonnegative continuous functions equal to
$\frac{1}{s}$ on $[\frac{1}{n},1]$ and equal to $n$ on
$[0,\frac{1}{n}]$ so that $$\lim_{n\to   \infty}\int_0^1 f_n(s)
sd\beta(s) = y_0.$$  Then,  
\begin{eqnarray}
\int_{\frac{1}{n}}^1 f_n(s) d\alpha(s) & \leq & \int_0^1 f_n(s)
sd\beta(s) .\label{est41}
\end{eqnarray}
Since the right hand side of (\ref{est41}) converges as $n \to
\infty,$ we conclude that (\ref{converge}) holds.  Since the right
hand side converges to $y_0,$ we conclude that $L = \int_0^1 s^{-1}
d\alpha(s)$ is a lower bound for any $y_0$ augmenting the original
sequence.  Since we have already established that when the integral
converges, $y_0=L$ gives a completely monotonic augmented sequence, we
are done. $\thmend$ 

{\sc Remark:}  With $\{y_i\}_{i=1}^\infty$ and $L$ as in Lemma
  \ref{extending}, any $y_0 \geq L$   gives a completely monotonic
  augmented sequence.

\begin{Lemma}\label{tableaudetermined} Let $X$ be a tableau.  Then
  $X$ is determined by its first column.  In fact, if $\{X_{n,0}\} =
  \{X_{n,0}\}_{n=0}^\infty $ is the first column of $X$ and 
  $\alpha:[0,1] \to \R$ is the normalized nondecreasing 
function representing $\{X_{n,0}\}:$   
\begin{eqnarray}
X_{n,0} & = & \int_0^1 s^n d\alpha(s), \label{col14}
\end{eqnarray} 
then 
\begin{eqnarray}
X_{0,p} & = & \int_0^1 s^{-p}(1-s)^p d\alpha(s). \label{row31}
\end{eqnarray}
\end{Lemma}
{\sc Proof} By Lemma \ref{columnsmonotonic} $\{X_{n,0}\} =
\{X_{n,0}\}_{n=0}^\infty$ is a completely monotonic sequence and thus
admits a representation by $\alpha$ as in (\ref{col14}).  By
definition of a tableau, $X_{n,0} +X_{n,1} = X_{n-1,0}$ for all $n\geq
1,$ and thus for $n \geq 1,$  
\begin{eqnarray*}
X_{n,1} & = & \int_0^1 s^{n-1}(1-s) d\alpha(s).
\end{eqnarray*}
Since $\{X_{n,1}\}_{n=1}^\infty$ is represented as a moment sequence,
by Hausdorff's Theorem $\{X_{n,1}\}_{n=1}^\infty$ is completely
monotonic.  Since $\{X_{n,1}\}_{n=1}^\infty$ is part of a column of a
tableau, $\{X_{n,1}\}_{n=1}^\infty$ extends to a completely monotonic
sequence $\{X_{n,1}\}_{n=0}^\infty .$  By Lemma \ref{extending}, 
we conclude that 
\begin{eqnarray*}
 &  & \int_0^1 s^{-1}(1-s) d\alpha(s)
\end{eqnarray*} 
converges and we set 
\begin{eqnarray*}
L & = & \int_0^1 s^{-1}(1-s) d\alpha(s).
\end{eqnarray*} 
Let $\beta:[0,1] \to \R$ be a normalized nondecreasing function
representing the completely monotone sequence
$L, X_{1,1}, X_{2,1}, \dots$  Let $\ep = X_{0,1} - L$ and let $h(t)$
be the Heaviside function:  
\begin{eqnarray*}
h(t) & = & \begin{cases} 
            1 & \hbox{ if } t > 0 \\
            0 & \hbox{ if } t \leq 0. \end{cases}
\end{eqnarray*}
Define $\gamma :[0,1] \to \R$ by
\begin{eqnarray}
\gamma(t)& = & \beta(t) + \ep h(t).
\end{eqnarray}
Then $\gamma $ is nondecreasing and for all $n \geq 0,$
\begin{eqnarray*}
X_{n,1} & = & \int_0^1 s^n d\gamma(s).
\end{eqnarray*}
Let $f_n(s)$ be as defined in the proof of Lemma \ref{extending}.
  Consider the pair of columns $\{X_{n,1}\}$ and $\{X_{n,2}\}.$  By
  the   analysis given for the pair $\{X_{n,0}\}$ and $\{X_{n,1}\},$
  we know   that $$\int_0^1 s^{-1}(1-s)d\gamma(s)$$converges and
  therefore $\int_0^1f_n(s) (1-s) d\gamma(s)$ converges as $n \to 
  \infty.$  But $\int_0^1 f_n(s) (1-s) d\beta(s)$ is nonnegative and
  $\int_0^1 f_n(s) (1-s) dh(s)$ diverges as $n \to \infty,$ 
  from which we conclude that $\ep = 0.$ 
Thus, $X_{0,1} = L$ and the column $\{X_{n,0}\}$ determines the column
  $\{X_{n,1}\}.$ The lemma follows by induction. $\thmend$

\hfill

{\sc Proof of Theorem \ref{maintheorem}} Suppose that $T \in \S$ and
suppose that $\alpha:\R^+ \to \R$ is a normalized
nondecreasing function representing $T:$ 
\begin{eqnarray*}
t_n & = & \int_0^\infty s^n d\alpha(s).
\end{eqnarray*}
Fix $p \in \N$ and define $\beta: \R^+ \to \R^+$ by 
\begin{eqnarray*}
\beta(s) & =& \int_0^s \frac{1}{(1+u)^p} d\alpha(u).
\end{eqnarray*}
Then $\beta $ is nondecreasing on $\R^+$ and of bounded variation.
Let $S$ be the sequence corresponding to the moments of $\beta:$ $$s_n
= 
\int_0^\infty u^n d\beta(u).$$  A direct computation using
(\ref{closure2}) gives $\rop^p(S) = T.$  This proves that every moment
sequence is stable. 

To establish the converse, suppose that $T$ is a stable sequence.  By
Lemma \ref{firstrowstable}, there is a tableau, $X,$ which
has $T$ as its first row.  By Lemma \ref{tableaudetermined}, $X$ is
determined by its first column.  By Lemma \ref{columnsmonotonic}, the first
column of $X$ is completely monotonic and thus, by Hausdorff's
Theorem, there is a unique normalized nondecreasing $\alpha:[0,1]\to
\R$ which represents $\{X_{n,0}\}:$ 
\begin{eqnarray*}
X_{n,0} & = & \int_0^1 s^nd\alpha(s).
\end{eqnarray*}
Thus, $T$ is determined by $\alpha.$  To complete the proof, we use
$\alpha $ to construct a measure on $\R^+$ with moments given by $T.$  

Write $u = \frac{1-s}{s}$ and $s = \frac{1}{1+u}.$  The function
$\gamma:\R^+\to \R$ defined as the composition $\gamma(u) = -\alpha(s)$
is nondecreasing with total variation bounded by the
variation of $\alpha.$  By Lemma \ref{tableaudetermined}, 
\begin{eqnarray}
X_{0,n} & = & \int_0^\infty u^n d\gamma(u) 
\end{eqnarray}
which exhibits the first row of $X$ as a moment sequence and completes
our proof. $\thmend$

\section{Applications}

Theorem \ref{maintheorem} provides for a representation of the
transition probabilities for a Markov chain which provides a classical
precursor for a discrete theory of quantum gravity satisfying
causality, covariance and cosmological renormalizability:

\begin{Corollary}\label{probabilities} Suppose that $M \in \m$ is a
  Markov chain satisfying Definition \ref{modelsofqg}.  Suppose that
  $T \in \s$ is the sequence of coupling constants defining $M$ and
  that $\alpha:\R^+ \to \R$ is a nondecreasing function representing
  $T.$  Suppose that $\{q_n\}_{n=0}^\infty$ are the transition
  probabilities defined in (\ref{qn}).  Then 
\begin{eqnarray}
\frac{1}{q_n} & = & \int_0^\infty (1+s)^n d\alpha(s). \label{transp15}
\end{eqnarray}
\end{Corollary}
{\sc Proof}  This follows immedately from (\ref{tkqk}) and the binomial
theorem. $\thmend$

Using Corollary \ref{probabilities} we obtain an attractive
representation for general transition probabilities:

\begin{Corollary}\label{transprobabilities} Suppose that $C \in \Cs_n$
  and that $D \in F(C).$  Suppose that the cardinality of
  $\precur(C,D)$ is $\rho$ and that the cardinality of $\max(C,D)$ is
  $m.$  Suppose that $M \in \m$ is a   Markov chain satisfying
  Definition \ref{modelsofqg}.  Suppose that $T \in \s$ is the
  sequence of coupling constants defining $M$ and   that $\alpha:\R^+
  \to \R$ is a nondecreasing function representing $T.$  Then
\begin{eqnarray}
\prob(C \to D) & = & \frac{ \int_0^\infty s^m(1+s)^{\rho-m}
  d\alpha(s)}{\int_0^\infty (1+s)^n d\alpha(s)} . \label{transp25}
\end{eqnarray}
\end{Corollary}

{\sc Proof}  This follows immediately from Corollary
\ref{probabilities}, (\ref{probcd}), and the binomial theorem applied
to $s^m = ((1+s) -1)^m.$  $\thmend$

Our next result establishes that {\em all}  positive
sequences which grow fast enough are stable: 

\begin{Corollary}\label{fastgrowing} Any monotonic sequence which
  grows sufficiently quickly defines an element of $\St.$  
\end{Corollary}

{\sc Proof} For a quickly growing sequence, the positivity condition
on the Hankel determinants (\ref{hankel1}), (\ref{hankel2}), are
trivially satisfied as the value of the determinant is controlled by
the entry in the lower right hand corner.  Thus, any monotonic
sequence which grows sufficiently quickly is a moment sequence and 
Corollary \ref{fastgrowing} follows from Theorem
\ref{maintheorem}. $\thmend$  

Corollary \ref{transprobabilities} and Corollary \ref{fastgrowing}
provide a means of quantifying the evolution of causets under dynamics
which provide for rapidly increasing coupling constants.  We hope to
return to this in a future paper.  

Our final result uses Hankel determinants to associate to any stable
sequence which is not a finite linear combination of percolation
sequences, a model Hilbert space and a nonnegative self-adjoint
operator.  Our development follows that of Simon \cite{Si}. 

The Hankel determinants appearing in (\ref{hankel1}) and
(\ref{hankel2}) are associated to quadratic forms which arise
naturally in the analysis of the Stieltjes Moment Problem.  More
precisely, given a sequence $T\in \St,$ consider the sesquilinear
forms $H^i_N: \C^N \to \C$ defined by 
\begin{eqnarray}
H^0_N(\rho,\sigma) & = & \sum_{0 \leq n,m \leq N-1}
\bar{\rho}_n\sigma_m t_{n+m} \label{form031} \\
H^1_N(\rho,\sigma) & = & \sum_{0 \leq n,m \leq N-1}
\bar{\rho}_n\sigma_m t_{n+m+1}. \label{form131} 
\end{eqnarray}
Let ${\mathcal H}^i_N$ be the matrices associated to the forms $H^i_N$
via the relation
\begin{eqnarray}
H^i_N(\rho,\sigma) & = & \pair{\rho}{{\mathcal H}^i_N\sigma}
\label{pairing31} 
\end{eqnarray}
where the pairing is Euclidean. Then the Hankel determinants appearing
in Theorem \ref{stieltjes} are given by $\det({\mathcal H}^i_N ) =
H_{i,N}$ and the forms $H^i_N$ are strictly positive definite if and
only if the corresponding Hankel determinants are positive (cf
\cite{Si}).  
Following \cite{Si}, we use this material to reformulate the Stieltjes
result in the language of self-adjoint operators.  

For the remainder of the paper we assume that the sequence $T$ is not
a finite linear combination of percolation sequences, so that the
Hankel determinants are all strictly positive definite. 

Suppose that $\C[x],$ is the algebra of polynomials with complex
coefficients.  Define a positive definite inner product on $\C[x]$ by  
\begin{eqnarray}
\pair{p}{q} & = & H_{0,N}(\rho,\sigma) \label{innerproduct31}
\end{eqnarray}
where $p(x) = \sum_{n=0}^{N-1} \rho_n x^n $ and $q(x) =
\sum_{n=0}^{N-1} \sigma_n x^n.$  Using this inner product, we complete
$\C[x]$ to a Hilbert space $\h_T,$ where the subscript $T$ denotes the
dependence on the moment sequence $T.$  

Let $A$ be the operator with domain $D(A)= \C[x] \subset \h_T$ defined
by 
\begin{eqnarray}
A(p)(x) & = & xp(x). \label{definitionofa}
\end{eqnarray}
Then $A$ is densely defined, symmetric and nonnegative.  Thus, by the
theory of von Neumann, $A$ admits self-adjoint extensions.  Amongst
the (possibly many) self-adjoint extensions of $A$ there is a
distinguished extension, the {\em minimal} nonnegative self-adjoint
extension (the Friedrichs extension) of $A$ to an  operator $A_F$ with
domain contained in $\h_T.$  

It is a theorem of Simon that the collection of such extensions of $A$
parameterizes solutions to the (nondegenerate) Stieltjes moment
problem \cite{Si}.  We summarize these results in the following  
\begin{Theorem}\label{simon}(Simon) Suppose that $T \in \S$ is a
  sequence which is not a finite linear combination of percolation
  sequences and whose corresponding Stieltjes problem admits
  a solution.  Let $\h_T$ be the corresponding Hilbert space completion
  of the algebra of polynomials with inner product defined by 
  (\ref{innerproduct31}), and let $A:D(A) \to \C[x]$ be the operator
  defined by (\ref{definitionofa}).  Then every solution to the
  Stieltjes problem for the sequence $T$ corresponds to a unique
  nonnegative self-adjoint extension of $A$ to an operator
  $\tilde{A}:\h_T \to \h_T$ with spectral measure $\mu_{\tilde{A}}$
  satisfying $$t_n = \int_0^\infty s^n d\mu_{\tilde{A}}(s).$$
\end{Theorem}
With the conventions established above, we have the following corollary:
\begin{Corollary}\label{representingwithA}To every sequence $T\in
  \S_1 \cap \St,$ which is not a finite linear combination of
  percolation sequences, there corresponds a pair $(\h_T, A_F)$ where
  $\h_T$ is the Hilbert space completion of $\C[x]$ defined by inner
  products (\ref{innerproduct31}) and $A_F$ is the minimal nonnegative 
  self-adjoint extension of the densely defined operator $A:\C[x] \to
  \C[x]$ defined in (\ref{definitionofa}).  Thus, there is a
  distinguished spectral measure whose moments are given by the
  sequence $T.$
\end{Corollary}


\begin{thebibliography}{MM1} 


%
%


\bibitem[1]{RS} D. Rideout and R. Sorkin  {\em A classical sequential
    growth dynamics for causal sets} Phys. Rev. D {\sl \bf 61} 024002
    (2000).  

\bibitem[2]{R} I. Raptis {\em Quantum Space-Time as a Quantum Causal
Set} Preprint ArXiv:gr-qc/0201004 (2002).

\bibitem[3]{So} R. Sorkin, {\em Indications of causal set cosmology},
Internat. J. Theort. Phys. {\sl \bf 39} 1731 (2000).

\bibitem[4]{BLMS} L. Bombelli, J. Lee, D. Meyer and R. Sorkin {\em
Space-time as a Causal Set}, Phys. Rev. Lett. {\sl \bf 59} 521
(1987). 

\bibitem[5]{RZ} I. Raptis and R. Zapatrin {\em Algebraic description
of spacetime foam} Class. Quantum Grav. {\sl \bf 18} 4187 (2001).

\bibitem[6]{RW} T. Regge and R. M. Williams  {\em Discrete structures
in gravity} Jour. Math. Phys. {\sl \bf 41} 3964 (2000).  

\bibitem[7]{MORS} X. Martin, D. O'Connor, D. Rideout and R. Sorkin
    {\em  ``Renormalization'' transformations induced by cycles
    of expansion and contraction in causal set cosmology,}
Phys. Rev. D (3), {\sl \bf 63} 084026 (2001). 

\bibitem[8]{A} N. Akhiezer  {\em The Classical Moment Problem}, 
    Hafner,  New York  (1965).

\bibitem[9]{W} D. V. Widder  {\em The Laplace Transform},
    Princeton University Press,  Princeton, NJ  (1941).

\bibitem[10]{Si} B. Simon {\em The classical moment problem as a
    self-adjoint finite difference operator,} Adv. in Math. {\sl \bf
    137}  82 (1998).

\end{thebibliography}
\end{document}